\documentstyle [12pt,epsfig,aasms4] {article}

\begin{document}
\newcommand{\be}{\begin{eqnarray}}
\newcommand{\ee}{\end{eqnarray}}
\newcommand{\msun}{~{\rm M}_\odot}
\def\lsim{\mathrel{\rlap{\lower4pt\hbox{\hskip1pt$\sim$}}
	\raise1pt\hbox{$<$}}} 
\def\gsim{\mathrel{\rlap{\lower4pt\hbox{\hskip1pt$\sim$}}
	\raise1pt\hbox{$>$}}} 
\def\ni{\noindent}
\def\ul{\underline}
\def\in{\indent}
\def\lin{\in\in\in}
\def\etal{{\it et al.} }
\def\eg{{\it e.g.,} }
\def\ie{{\it i.e.,} }

\title{ CONTRIBUTION OF HIGH--MASS BLACK HOLES TO MERGERS OF COMPACT BINARIES }
\author{ HANS A. BETHE }
\affil{Floyd R. Newman Laboratory of Nuclear Studies \\
        Cornell University \\
        Ithaca, New York 14853, USA}
\author{ G. E. BROWN}
\affil{Department of Physics and Astronomy \\
        State University of New York\\
        Stony Brook, New York 11794-3800, USA}

\begin{abstract}
We consider the merging of compact binaries consisting of a high-mass
black-hole and a neutron star. From stellar evolutionary
calculations which include mass loss we estimate that a ZAMS mass
of $\gsim 80\msun$ is necessary before a high-mass black hole can result
from a massive O-star progenitor.

We first consider how Cyg X-1 with its measured orbital
radius of $\sim 17 R_\odot$
might evolve. Although this radius is substantially less than the initial
distance of two O-stars, it is still so large that the resulting compact
objects will merge only if an eccentricity close to unity results from
a high kick velocity of the neutron star in the final supernova explosion.
We estimate the probability of the necessary
eccentricity to be $\sim 1\%$; i.e., $99\%$ of the time the explosion of
a Cyg X-1 type object will end as a binary of compact stars which will not
merge in a Hubble time (unless the orbit is tightened in common envelope
evolution which we discuss later).

Although we predict $\sim 7$ massive binaries of Cyg X-1 type, we argue
that only Cyg X-1 is narrow enough to be observed and that only it has
an appreciable chance of merging in a Hubble time. This gives us a
merging rate of $\sim 3\times 10^{-8} {\rm yr}^{-1}$ in the galaxy,
the order of magnitude of the merging rate found by computer driven
population syntheses, if extrapolated to our mass limit of $80\msun$
ZAMS mass for high-mass black hole formation.
Furthermore, in both our calculation and in those of population
syntheses, almost all of the mergings involve an eccentricity
close to unity in the final explosion of the O-star.
>From this first part of our development we obtain only a negligible
contribution to our final results for mergers, and it turns out to be
irrelevant for our final results.

In our main development, instead of relying on observed binaries,
we consider the general evolution of binaries of massive stars. The critical
stage is when the more massive star A has become a black hole and the less
massive star B is a giant, reaching out to A. We then have a common envelope,
and we expect hypercritical accretion to A. A will accept a small fraction
of the mass of the envelope of B but will plunge deep into B while expelling
B's envelope.

We expect that star B can at least be in the mass range $15 \sim 35 \msun$
while the black hole A has a mass of $10\msun$. About 20 percent of the
binaries of this type
are found to end up in a range of orbital radii
favorable for merging; i.e., outside of
the relevant Roche Lobes, but close enough so that these final binaries
of compact objects will merge in a Hubble time.   The narrow
black-hole, O-star orbits do not seem to be found in population
syntheses because in them mergers happen almost completely as a result of
kick velocities.
In the exception, Case H of Portegies Zwart \& Yungelson (1998) which
includes hypercritical accretion, common envelope evolution is more effective
and we are in agreement with their results.

We find that the high-mass black-hole, neutron-star systems contribute
substantially to the predicted
observational frequency of gravitational waves.

We discuss how our high mass for high-mass black hole formation can be
reconciled with the requirements of nucleosynthesis and indicate that a bimodal
distribution of masses of black holes in single stars can account, at least
qualitatively, for the many transient sources which contain high-mass black
holes.
\end{abstract}

\section{ INTRODUCTION}
The supernova community used to believe that stars above a certain mass, about
$30-40 \msun$ ZAMS (zero age main sequence)
will collapse into a massive black hole
(MBH) of mass of order $10\msun$.  The argument was that in these stars the
mantle was bound with a binding energy well above $10^{51}$ erg so that the
supernova shock was not strong enough to expel it.

Whereas this may be true for single stars, Woosley, Langer \& Weaver
(WLW,1995)
showed that in binaries, where the hydrogen envelope of the primary star has
been transferred to the companion in RLOF (Roche Lobe Overflow), the evolution
of the resulting ``naked" He star; \ie the star without hydrogen envelope, led
to a substantially smaller presupernova core than that of a single star with
hydrogen envelope.  A comparison of compact core masses from naked He stars and
those evolved by Woosley \& Weaver (1995) for single stars is shown in
Figure~1, taken from
Brown, Weingartner \& Wijers (1996).  Detailed reasons for the great difference
in the evolution of ``clothed" and naked He cores are given in WLW (1995).

\begin{figure}
\centerline{\epsfig{file=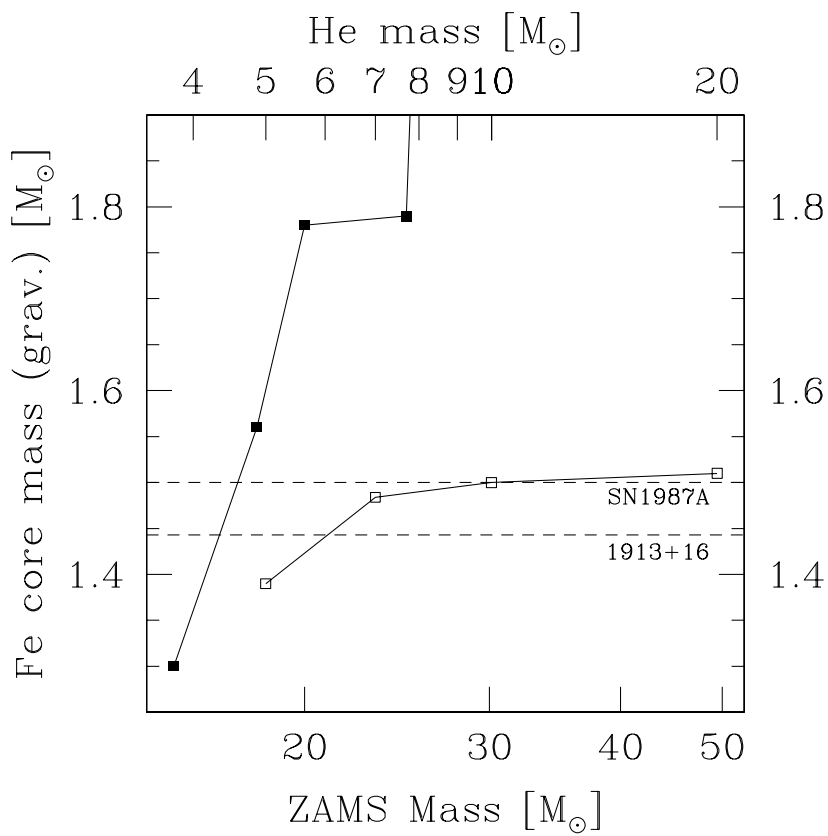,height=11cm}}
\caption{Comparison of the compact core masses resulting from the
evolution of single stars (filled symbols), Case B of Woosley \& Weaver (1995),
and naked helium stars (Woosley, Langer \& Weaver, 1995) with masses equal to
the corresponding He core mass of single stars.  The horizontal dashed lines
indicate the mass of the heaviest known well--measured pulsar and the maximum
mass of a neutron star.}
\label{fig1}
\end{figure}

Stars with ZAMS masses $\gsim 40\msun$ lose their masses by strong winds,
whether in binaries or not, and become Wolf--Rayet stars.  In an earlier paper
WLW (1993) investigated ZAMS masses of 35, 40, 60 and $85\msun$.  In those up
through $60 \msun$ the hydrogen envelope was blown off early enough for the He
cores to evolve as naked ones and compact core masses were around $1.5 \msun$
(gravitational).
In fact, with inclusion of extensive mass loss the lower line in
Fig.~\ref{fig1}, which heads just above $1.5 \msun$ for the higher
ZAMS masses, gave the WLW (1993) correspondence of Fe core mass to ZAMS
mass for single stars of masses $35-60 \msun$. Thus, rapid mass loss by
wind in this mass region which removes the H-envelope before
appreciable He core burning begins leaves a He core which burns
as a ``naked" one.  (In the case of the massive
stars the relation shown in Fig.~\ref{fig1} between Fe core mass and
He core mass no longer holds because of large wind losses in the latter.)
In the case of the $85\msun$ star some hydrogen envelope
remained during an appreciable part of the He burning,
so the He core burned as a (partially) clothed one
 and the compact core was more massive,
in the range $1.7 - 2.0\msun$, depending upon $^{12}C(\alpha,\gamma)^{16}O$
burning rate. For the very massive stars, the He core burns, at least part of
the time, as if clothed.  The mass loss rate used by WLW (1993) has been shown
to be too high.  Such rates were obtained from the free--free fluxes and
modelling of infrared spectral lines which involve the density quadratically.
Measurements from the polarization of the radiation (St-Louis et al. 1993;
Moffat \& Carmelle 1994) obtain a mass loss rate of $\lsim$ 50\% of that used
by WLW.  The polarization depends linearly on the density, provided that the
wind is optically thin to electron scattering.  Thus, estimates based on this
are independent of the inhomogeneities which are known to be present.  The
lower mass loss rates come into agreement with the $\dot{M}$ from dynamical
arguments.  Certainly the WLW calculations should be redone with the lower
rates, which may change our conclusions.

\setcounter{equation}{0}
\section{THE COMPACT STAR}
\label{sec:2}

Thorsson, Prakash \& Lattimer (1994) and Brown \& Bethe (1994) have
studied the compact core after the collapse of a supernova, assuming
reasonable interactions between hadrons. Initially, the core consists
of neutrons, protrons and electrons and a few neutrinos.
It has been called a proto-neutron star. It is stabilized against
gravity by the pressure of the Fermi gases of nucleons and leptons,
provided its mass is less than a limiting mass $M_{PC}$
(proto-compact) of $\sim 1.8 \msun$.

If the assembled core mass is greater than $M_{PC}$ there is no
stability and no bounce; the core collapses immediately into a black hole.
It is reasonable to take the core mass to be equal to the mass of the
Fe core in the pre-supernova, and we shall make this assumption,
although small corrections for fallback in the later supernova explosion
can be made as in Brown, Weingartner \& Wijers (1996). If the center collapses
into a black hole, the outer part of the star has no support and will also
collapse. We then get a massive black hole containing the entire mass of the
presupernova star, perhaps of order $\lsim 10\msun$. (At least in binaries
the companion star will have removed the hydrogen envelope by either
Roche Lobe Overflow or by common envelope evolution, depending on the
mass of the companion, and wind will diminish the He core before the
SN explosion.)

If the mass of the core is less than $M_{PC}$, the electrons will be
captured by protons
\be
p+e^- \rightarrow n+\nu
\ee
and the neutrinos will diffuse out of the core. This process takes of order
of 10 seconds, as has been shown by the duration of the neutrino signal
from SN1987A. The result is a neutron star, with a small concentration of
protons and electrons.
The Fermi pressures of the core are chiefly from the nucleons, with small
correction from the electron. On the other hand the nucleon energy is
increased by the symmetry energy;
i.e., by the fact that we now have nearly pure neutrons instead of
an approximately equal number of neutrons and protons.
Throsson et al. (1994) have calculated that the maximum mass of the
neutron star $M_{NS}$ is still about $1.8\msun$; i.e., the symmetry
energy compensates the loss of the Fermi energy of the leptons.
Corrections for thermal pressure are small (Prakash et al. 1997).

The important fact is that the ten seconds of neutrino diffusion from
the core give ample time for the development of a shock which expels
most of the mass of the progenitor star.

But this is not the end of the story. The neutrons can convert into
protons plus $K^-$ mesons,
\be
n\rightarrow p+K^-.
\label{n2.2}
\ee
Since the density at the center of the neutron star is very high, the
energy of the $K^-$ is very low, as confirmed by Li, Lee \& Brown (1997)
using experimental data. By this conversion the nucleons can again become
half neutrons and half protons, thereby saving the symmetry energy needed
for pure neutron matter. The $K^-$, which are bosons, will condense,
saving the kinetic energy of the electrons they replace. The reaction
(\ref{n2.2}) will be slow, since it is preceded by
\be
e^-\rightarrow K^- +\nu
\label{n2.3}
\ee
which is actually effected in the star by reactions such as
$e^-+p\rightarrow K^-+p+\nu$, with strangeness breaking. (Times are long
enough for chemical equilibrium to be realized.)
It becomes energetically
advantageous to replace the fermionic electrons by the bosonic $K^-$'s
at higher densities.
Initially the neutrino states in the neutron star are filled up to
the neutrino chemical potential with trapped neutrinos, and it takes some
seconds for them leave the star.
These must leave before new neutrinos can be formed from the process
(\ref{n2.3}). Thorsson et al. (1994) have calculated that the maximum
mass of a star in which reaction (\ref{n2.2}) has gone to completion is
\be
M_{NP} \simeq 1.5\msun\ ,
\ee
where the lower suffix $NP$ denotes their nearly equal content of
neutrons  and protons, although we continue to use the usual name
``neutron star".
This is the maximum mass of neutron stars, which is to be compared with
the masses determined in binaries. The masses of 19 neutron stars
determined in radio pulsars (Thorsett \& Chakrabarty, 1998) are
consistent with this maximum mass.

The core mass $M_C$ formed by the collapse of supernova must therefore be
compared to the two limiting masses, $M_{PC}$ and $M_{NP}$. If
\be
(I) \;\; M_C > M_{PC}
\ee
we get a high mass black hole, of mass essentially equal to the full mass
of the presupernova star. If
\be
(II) \;\; M_{PC} > M_C > M_{NP}
\ee
we get a low mass black hole, of mass $M_C$. Only if
\be
(III) \;\; M_C < M_{NP}
\ee
do we get a neutron (more precisely, ``nucleon") star from the SN.
Only in this case can we observe a pulsar. In cases (II) and (III)
we can see a supernova display. In case (I) only initial neutrinos
from electrons captured in the collapse before $M_C$ becomes greater than
$M_{PC}$ can be observed but no light would reach us.


We tentatively choose the lower limit of ZAMS mass for making MBHs to be
$80\msun$.  On the other hand it is believed that ZAMS above $100\msun$ do not
exist, because of excessive formation of electron pairs.  So we assume that the
range of 80 to $100\msun$ is available.

\setcounter{equation}{0}
\section{RATE OF FORMATION}
\label{sec:3}

We are interested in massive binaries containing one star of ZAMS mass
between 80 and
$100\msun$.  As in Bethe \& Brown (1998) we start from the assumption that
there is one supernova per century per galaxy in a binary.  Assuming also that
$10\msun$ is required for a star to end up as a SN of type II (or Ib or c),
the
formation of binaries of $M > 10\msun$ is also $10^{-2}$ per year
per galaxy.  We assume a Salpeter function with index $n = 1.5$; then the
fraction of such stars between 80 and $100\msun$ is
\be
f = 8^{-3/2} - 10^{-3/2} = 1.26 \times 10^{-2} \ .
\ee
So the rate of formation is
\be
\alpha_1 = 1.26 \times 10^{-4} {\rm yr}^{-1} {\rm per~galaxy} \ .
\label{eq:3.2}
\ee

We require that this star $A$ be accompanied by a companion $B$ of $M >
10\msun$.  Assuming that the distribution of mass of the companion is $dq$,
with
\be
q = M_B/M_A
\ee
and having assumed $M_A = 90\msun$ on average, we need $q > 1/9$ which has a
probability
\be
1 - q = 0.9 \ ,
\ee
hence formation rate
\be
 \alpha_2 = (1-q)\alpha_1 = 1.13 \times 10^{-4} {\rm yr}^{-1} \ .
\ee

In order to observe strong X--rays from the MBH, the distance $a$ between star
$A$, the MBH, and star $B$, an $O$ or $B$ star, should not be too large, let us
say $< 150~{\rm R}_\odot$.  On the other hand it must not be
$< 30~{\rm R}_\odot$
because otherwise the two stars would merge already at this stage of
evolution.  Assuming, as in Bethe \& Brown (1998) a distribution $da/7a$, the
probability that $a$ falls in the desired limits, is
\be
p = 7^{-1} \ln (150/30) = 0.23
\label{eq:3.6}
\ee
giving for the probability of formation (per galaxy)
\be
\alpha_3 = 0.23 \alpha_2 = 2.6 \times 10^{-5} {\rm yr}^{-1} \ .
\label{eq:3.7}
\ee
Although we take the same logarithmic distribution as in Bethe \& Brown, we
think of the lower limit as $a > 2\times 10^7$ km so that the more massive
stars we deal with here lie inside their Roche Lobes and the upper
limit as $a<2\times 10^{10}$ km, although the latter is uncertain because
O-star
binaries probably will not be recognized for such a large separation
(Garmany et al. 1980).
High mass transfers from the $O$--star occur only when it nearly fills its
Roche Lobe.  It is then bright for a time (Massevich \etal, 1979)
\be
\tau = 2.7 \times 10^5 \; {\rm yr} \ .
\label{eq:3.8}
\ee
so the expected number of strong X--ray emitting binaries in the galaxy is
\be
N = \alpha_3 \tau = 7 \ .
\label{eq:3.9}
\ee

Only one such binary has been observed, Cygnus X--1, at a distance of 2.5 kpc.
It might be thought that X-ray binaries at larger distance have escaped
detection because of absorption of X-rays in the highly ionized galactic
medium. But
in fact, X-rays of energy $>3 keV$ should be seen throughout the galaxy,
penetrating even the Galactic disc. On the other hand, we are probably
seeing the closest binary of Cyg X-1 type, with orbital separation
only about half of the initial separation in the double O-star progenitor.
Thus, other such objects may well have substantially lower luminosities.
The O-star in such a binary would probably be in the disc and might well
not be observed at greater distances, so that the X-rays could not be
associated with the binary. In any case, it is clear that in the disc there
is only one bright Cyg X-1 type object, a fact we use below.

In the Magellanic clouds with less than one-tenth the mass of the galaxy,
two high-mass black-hole binaries, LMC X-3 with a B-star companion and
LMC X-1 where the donor is probably an O-star, have been observed.
We believe the high incidence of HMBHs in the LMC to be not only a
consequence of the large amount of star formation there, but also of the
low metallicity compared with solar. We return to a discussion of high-mass
black hole formation in Section \ref{sec:8}.

\setcounter{equation}{0}
\section{MERGERS}
\label{sec:4}

In the case of low-mass black holes (LBH), the rate of mergers is limited
chiefly by the disruption of the binary due to the recoil that star B
experiences when it goes supernova. For massive black holes (MBH)
this is of little concern since the orbital velocity of the binary
before the SN process is already of order 600 km s$^{-1}$ (see below),
comparable to the higher recoil velocities, so that the recoil is
unlikely to distrupt the binary.

Instead of this, the concern for MBH is the two stars
of the binary may be too far apart so that they will not merge during
Hubble time (assumed to be $10^{10}$ years).
Note that the progenitor binary must have separation $a > 30 R_\odot$
in order that the two O-stars not fill their Roche Lobes.
At present, the separation in Cyg X-1 is only slightly more than half
this,
which is permissible because now only the O-star needs to be in its
Roche Lobe, the black hole has negligible radius. But as we show below, even
this small separation is
still large enough to substantially cut down
the probability of merging within a Hubble time. Because of rapid mass
loss in such massive stars of ZAMS $80-100\msun$ there is no well
established procedure for calculating their evolution in binaries.
Starting from a separation $>30\msun$ mass exchange by the very
massive star
with the lower mass companion when the former is on the main sequence
(Case A mass transfer) could decrease the separation substantially, perhaps
causing the stars to coalesce.
However, the resulting binary would then widen (roughly back  to
its original separation) as the Wolf-Rayet, which remains after mass
transfer from the very massive star, loses mass by wind.
We do not have any evidence of the net change
(S. Portegies Zwart, private communication).

Because of the
uncertainties in the evolutionary scenario, we adopt an empirical
approach, beginning from the measured separation $a_i$ in Cyg X-1.
The requirement that the two stars merge in a Hubble time then sets a lower
limit on the eccentricity $e$ of the final neutron-star black-hole orbit:
\be
\alpha_2 (1-e^2) < a_0
\label{eq:4.1}
\ee
where $a_2$ is the semi-major axis of the orbit after the SN event
and $e$ is the eccentricity of the orbit; $a_0$ will be calculated
in Section \ref{sec:5}.

The left hand side of (\ref{eq:4.1}) is related to the angular
momentum $J$ of the post-SN orbit. In fact
\be
J =a_2 V_3 (1-e^2)^{1/2}
\label{eq:4.2}
\ee
where $V_3$ is the orbital velocity if the orbit were circular.
(We do not need to consider the mass of the system.)
By Newton's laws,
\be
a_2 V_3^2 = G(M_A+M_n)
\label{eq:4.3}
\ee
where $M_A$ is the mass of the MBH and $M_n$ that of the neutron
star resulting from the SN of star B. Hence
\be
J^2 = G(M_A+M_n) a_2 (1-e^2) \ .
\label{eq:4.4}
\ee
On the other hand, immediately after the SN, the distance between stars
$A$ and $B$ is still $a_1$, and the relative velocity is
\be
\vec V_2 =\vec V_1 +\vec Q
\label{eq:4.5}
\ee
where $\vec Q$ is the recoil velocity due to the SN. The angular momentum is
\be
\vec J &=& \vec a_1 \times \vec V_2 \nonumber\\
J^2 &=& a_1^2 V_2^2 \sin^2 \psi
\label{eq:4.6}
\ee
where $\psi$ is the angle between $\vec V_2$ and $\vec a_1$.

This angle can be calculated but is complicated.
We now make the approximation of replacing $\sin^2\psi$ by
\be
\langle \sin^2\psi\rangle =\frac 23\ .
\label{eq:4.7}
\ee
Then, equating (\ref{eq:4.4}) and (\ref{eq:4.6})
and using (\ref{eq:4.1})
\be
a_1^2 \; V_2^2 < \frac 32 G (M_A+M_n)\;  a_0 \ .
\label{eq:4.8}
\ee
Eq. (\ref{eq:4.7}) involves a double approximation;
(1) $\psi$ is uniformly distributed over the sphere, and
(2) only the average of $\sin^2\psi$ is used, instead of
calculating the limit on $a_1 V_2$ separately for each possible
direction of $\vec Q$ and then averaging the results. We believe
that a correct calculation would give a somewhat higher
limit on $a_1 V_2$.

The product $a_1 V_2^2$ can be calculated assuming that the orbit
of the binary is circular before the SN event, an assumption which is
probably very nearly correct. Then, similar to (\ref{eq:4.3}),
\be
a_1 V_1^2 = G (M_A+M_B)
\label{eq:4.9}
\ee
where $M_B$ is the mass of star B, the companion of the MBH, before
the SN event. In Cyg X-1, it is believed that the mass of the MBH
is $M_A^\prime =10\msun$ and that of the O-star companion
$M_B^\prime = 17\msun$ (Herrero et al. 1995).
After RLOF of the O-star,\footnote{Later we shall show that common
envelope evolution of the black hole in the evolving O-star is more likely.
}
these masses happen to be just exchanged,
assuming conservative mass transfer,
so that immediately before the SN,
\be
M_A = 17\msun, \;\;\; M_B = 10\msun\ .
\label{eq:4.10}
\ee
However, we shall use ``standard"values
\be
M_A=M_B=10\msun .
\label{eq:4.10a}
\ee
The SN changes B to a neutron star, $M_n=1.4 \msun$,
while A remains unchanged. We have now
\be
a_1^2\;  V_1^2 &=& G (M_A +M_B)\;  a_1 .
\label{eq:4.11}
\ee
Dividing (\ref{eq:4.8}) by (\ref{eq:4.11})
\be
\frac{V_2^2}{V_1^2} < \frac 32 \left( \frac{M_A+M_n}{M_A+M_B}\right)
 \frac{a_0}{a_1} \equiv B .
\label{eq:4.12}
\ee
Usually, $a_1> a_0$ so $V_2 < V_1$.
>From (\ref{eq:4.5}) this means that the recoil
velocity $\vec Q$ of the SN must be approximately opposite
to the pre-SN orbital velocity $\vec V_1$, and of the same order
of magnitude.
Setting
\be
\vec Q\cdot \vec V_1 &=& -Q V\mu
\label{eq:4.13} \\
Q/V_1 &=&\omega
\label{eq:4.14}
\ee
thus (\ref{eq:4.12}) become
\be
1+\omega^2 -2\omega\mu < B
\label{eq:4.15}\\
1-\mu < \frac{B-(1-\omega)^2}{2\omega} \ .
\label{eq:4.16}
\ee
Since B is in general considerably smaller than 1, we thus get the
condition
\be
|1-\omega | < B^{1/2}
\label{eq:4.17}
\ee
stating once more that $Q$ must be of the same order as $V_1$.
The fraction of the total solid angle available to $\vec Q$ is
$(1-\mu)/2$.

Now according to (\ref{eq:4.11}),
\be
V_1^2 &=& \frac 23 \times \frac{10^{-7}\times 20\times 2\times 10^{33}}{a_1}
 = 2.7\times 10^{15}\times \frac{10^{12}}{a_1}
\label{eq:4.18} \\
V_1 &=& 520 {\rm km \; s^{-1}} (10^{12}/a_1)^{1/2} .
\label{eq:4.19}
\ee
In Cygnus X-1.
the distance between the two stars is
\be
a_1 = 17 R_\odot = 1.2\times 10^{12} {\rm cm} .
\label{eq:4.20}
\ee
So $a_1$ is likely to be of order $10^{12}$. This means
$V_1$ is of order 500 km s$^{-1}$, and $Q$ is likewise.
Now Cordes \& Chernoff (1997) have found that the recoil
velocities of pulsars are distributed bimodially with $\sigma=175$
and $700$ km s$^{-1}$, respectively. The smaller velocity
Gaussian is negligible for our $Q$. For the larger $\sigma$, we set
\be
x\equiv Q/\sigma =1 .
\label{eq:4.27}
\ee
The Gaussian distribution is
\be
f \; dx = 2\pi^{-1/2} x^2 dx e^{-x^2} .
\label{eq:4.28}
\ee
We assume $V_1=\sigma$ for the Gaussian with the greater $\sigma$,
then for this Gaussian, $\omega=x$ and the probability is,
cf. (\ref{eq:4.16})
\be
\int\frac{1}{4x} [B-(1-x)^2] f\; dx
\approx \frac{2 e^{-1}}{3\sqrt{\pi}}\; B^{3/2}
= 0.14\; B^{3/2} .
\label{eq:4.29}
\ee
For large $B$, substitute $0.9 B$ for $B^{3/2}$. Considering that
only $20\%$ of pulsar recoils are in the larger $\sigma$ Gaussian,
the fraction of binaries containing massive black holes which merge
during a Hubble time is
\be
\phi = 0.03\; B^{3/2}
\ee
with $B$ given by (\ref{eq:4.12}). Inserting our values for
$M_A$, $M_B$ and $M_n$,
\be
B= 0.85\; a_0/a_1  .
\label{eq:4.31}
\ee
It remains to calculate $a_0$.

\setcounter{equation}{0}
\section{MERGERS, NUMERICAL}
\label{sec:5}

If the orbit is circular, merger will take place within a Hubble time
if the initial distance is $<R_{max}$, with $R_{max}$ given by
Bethe \& Brown (1998), eq. (7.7)
\be
R_{max}^4= 1.6\times 10^{25}\; {\rm km}^4\; \frac{M_A M_B (M_A+M_B)}
{M_\odot^3} .
\label{eq:5.1}
\ee
$R$ is the same as the semi-major axis introduced in
Section \ref{sec:4}, so $a_0=R_{max}$. If the orbit is excentric, the merger
time for a given major axis is diminished by a factor (P. Eggleton, 1998,
private communication)
\be
Z(e) = (1-e^2)^{3.689-0.243 e- 0.058 e^2} .
\label{eq:5.2}
\ee
Therefore the condition on the initial semi-major axis, $a_2$, is
\be
 a_2^4\; Z(e) < R_{max}^4 .
\label{eq:5.3}
\ee
Taking the fourth root
\be
 a_2\; Z^{1/4} < R_{max}
\label{eq:5.4}
\ee
$Z^{1/4}$ is almost $1-e^2$, so the left hand side of (\ref{eq:5.4})
is almost $a_2 (1-e^2)$, thus
\be
a_2 (1-e^2) < R_{max} (1-e^2)^{0.0777+0.0608 e+ 0.0195 e^2} .
\label{eq:5.5}
\ee
The last factor depends only slightly on the eccentricity. So
we replace in it $e^2$ by an average of 0.5. This makes that factor
\be
0.5^{0.130}= 0.914 .
\label{eq:5.6}
\ee
Inserting in (\ref{eq:5.1}) $M_A=10 \msun$, $M_B=1.4\msun$,
\be
R_{max} = 7.1\times 10^6\; {\rm km} .
\label{eq:5.7}
\ee
The upper limit of $a_2(1-e^2)$ is the definition of $a_0$, according to
(\ref{eq:4.1}). Thus the right hand side of (\ref{eq:5.5}) equals $a_0$ ,
and we have
\be
a_0=6.5\times 10^6\; {\rm km}
\ee
and (\ref{eq:4.12}) becomes
\be
B=5.1\times 10^6\;{\rm km}/a_1 .
\label{eq:5.8}
\ee
The measured distance between the two stars in Cygnus X-1 is
$1.2\times 10^7$ km. If this distance does not change in RLOF, then
$a_1=1.2\times 10^7$ km,
\be
B &=& 0.42 \\
\phi &=& 0.009 .
\label{eq:5.10}
\ee
This estimate can be checked from Fig. 4a of Kalogera (1996). In our situation
her parameters are $\beta\approx 0.6$, $\xi\approx 0.9$.
Using the Eggleton formula eq. (\ref{eq:5.2}), we find that
$e\gsim 0.76$ for merger. The average survival probability is $\sim 0.3$
for these eccentricities
and our $\xi$ is $\sim 0.3$, so that the final probability is
$1/5\times 0.24\times 0.3\simeq 0.014$ close to our $\phi$.

Multiplying this by the formation rate $\alpha_3$, eq.(\ref{eq:3.7}),
we find for the rate of mergers
\be
\alpha_3\phi = 2\times 10^{-7} {\rm yr}^{-1}  .
\label{eq:5.11}
\ee
So this is a rare event as we might expect. In fact, (\ref{eq:5.11})
is probably still an overestimate because we predicted $7$ X-ray
emitting binaries in the galaxy (eq.(\ref{eq:3.9})) and we see only
Cyg X-1. In our assumed distribution of these objects out to $150 R_\odot$,
the others would be expected to have substantially larger $a_1$'s than
Cyg X-1, in which case their merger rate would be negligible.
Thus, a possible guess of the rate of mergers is not much larger than
$1/7$ of (\ref{eq:5.11}).

We compare our result with results  from population synthesis:
\begin{enumerate}
\item First we note that the smallness of the rate of mergers
results from the separation at the black-hole, O-star stage being
too large for merger (for spherical orbits) without a high
eccentricity from kick velocities.
Whereas Portegies Zwart \& Yungelson (1998) obtain a
(high-mass) black-hole, neutron-star birthrate of $18.7\times 10^{-5}$
yr$^{-1}$ in the galaxy without kick velocities, their merger rate
is zero for this case. Lipunov et al. (1997) find an order of
magnitude greater merger rate with inclusion of kick velocities than
without.

\item Portegies Zwart \& Yungelson \footnote{We compare with their Case H
in which hypercritical accretion was included in the next section.}
 use $40\msun$ for their ZAMS mass for
black hole limit and we compare with those for $40\msun$ from Lipunov
et al.. Increasing the mass limit to $80\msun$ and taking into account
the factor $1-q$ would decrease the results of these authors by
a factor of $\sim 5$. Thus, Portegies Zwart \& Yungelson would
have a merging rate of $2\times 10^{-7}$ yr$^{-1}$ and Lipunov et al.,
one of $\sim 4\times 10^{-8}$ yr$^{-1}$ for our mass limit of $80\msun$.
Our rate obtained by taking $1/7$ of (\ref{eq:5.11}) is
\be
R=3\times 10^{-8}\; {\rm yr}^{-1}
\label{eq:5.12}
\ee
about
the same as the latter. We believe that our simple considerations give
same understanding of the results obtained in the latter population
synthesis, and we shall return to a discussion of the work of
Portegies Zwart \& Yungelson later.
\end{enumerate}

We can also try an "observational estimate" of the merging of Cyg X-1
type objects, although we have only one such object near its
Roche Lobe. Assuming as in (\ref{eq:3.8}) Cyg X-1
to be bright for a time $\tau=2.7\times 10^5$ yr,
the birth rate of such bright objects
\be
\tau^{-1} = 3.7\times 10^{-6}\; {\rm yrs}^{-1} .
\label{eq:5.13}
\ee
Given the merging probability of the resulting black-hole, neutron-star
binary $\phi=0.009$ from (\ref{eq:5.10}) we arrive at a merging rate
of $4\times 10^{-8}$ yr$^{-1}$ for the galaxy, in rough agreement with
(\ref{eq:5.12}).
This may be an underestimate since Cyg X-1 will probably go into
stable mass transfer after losing substantial mass to the black
hole. LMC X-3 with roughly equal black hole and B-star masses
(Kuiper et al. 1997),
may already be in stable mass transfer. Thus the $2.7\times 10^5$ yrs
may well be an underestimate.

Although the last two sections give results for mergers which are negligible
compared to our final results in \S 7, once hypercritical accretion is included
in our common envelope evolution, we believe that they are instructive in that
they match rather well the results of population synthesis, especially those of
Lipunov et al.  In Portegies Zwart \& Yungelson (1998) (generally unstable)
mass transfer starts on the dynamical time scale of the donor, changing along
the way to the thermal time scale and the last part of the mass is sometimes
even transferred on a nuclear time scales (Portegies Zwart, private
communication).
Their initial separations of the O-star binaries are nearly double ours.
(We take ours before substantial main-sequence burning which increases
the stellar radius by a
factor $\sim$ 2.) Thus, their common envelope evolution does tighten
the orbits, but clearly not enough for substantial
merger of the compact objects evolved
later. The exception is Case H of Portegies Zwart \& Yungelson, the case
similar to ours with inclusion of hypercritical accretion, which gives a
merger rate $\sim 35$ times that of Cases B-G. We develop in the
next section why bypercritical accretion is so effective.

\setcounter{equation}{0}
\section{COMMON ENVELOPE EVOLUTION}
\label{sec:6}

In the last two sections we have addressed the evolution of massive
binaries in which the primary evolves into a high-mass black hole
in an empirical way. We match rather well the population synthesis results of
Lipunov et al. (1997). In particular, these authors find that introduction
of kick velocities increases their merging rate by an order of magnitude.
Portegies Zwart \& Yungelson (1998) find zero mergers without kick
velocities ! We believe this to result from the fact that their black-hole
O-star binaries lie too far apart for the resulting binary of compact objects
to merge in a Hubble time without the increase in merger distance given by an
eccentricity close to unity; i.e., these authors have very little (no) merging
from circular black-hole neutron-star orbits, except for their Case H which
includes hypercritical accretion which we discuss below.

Clearly the population syntheses of Lipunov
et al. (1997) do not include efficient common envelope evolution which
bring the compact object (high-mass black hole) close to the He
core of the companion star.

We now develop our scenario for common envelope evolution, which hypercritical
accretion makes particularly plausible, in that we have quantitative
control over the crucial quantities. As the envelope of the giant expands
to meet the compact object one can
see from Rasio \& Livio (1996) that the common envelope evolution begins
rather quickly.  Soon after the beginning of mass transfer the compact object
creates a tidal bulge in the evolving companion, transferring angular
momentum to the companion. As the giant companion
loses mass, the isentropic envelope responds by expanding.  The compact object
plunges into the companion, the chief loss in orbital energy occuring in
$\lsim 1$ year in time (Terman, Taam \& Hernquist 1995;
Rasio \& Livio 1996).  Our analytical estimate is roughly consistent with this
because our hypercritical accretion proceeds at $\sim 10^8$ Eddington, or about
$1\msun$ yr$^{-7}$ and the total accretion is of the order of solar masses,
giving the time in years.

In the literature common envelope evolution is thought to occur when the
two stars involved do not differ too much in mass, by not more than a factor
of 2 to 3. This condition arises because, as higher mass ratios,
 the mass receiving star will then
not be able to accept the large amount of mass transferred to it on the
short thermal time scale of the companion. This thermal time scale is
(van den Heuvel 1994):
\be
\tau_{th} =\frac{GM^2}{RL}\simeq
(3\times 10^7)/ (M/M_\odot)^2 .
\label{eq:6.1})
\ee
In fact, the factor 2 to 3 in $q$ initially came from a factor $\sim 10$ in
$\tau_{th}$, from the work of Kippenhahn \& Meyer (1977) who
considered Case A mass transfer (during main sequence evolution).
In Case B mass transfer the transfer takes place as the giant traverses
the Herzsprung gap, much more quickly than in Case A, so it is not
clear why $q$ could not be substantially reduced and still have
common envelope evolution.

The situation with hypercritical accretion is different from either
Case A or Case B (RLOF) mass transfer. Initially our $\sim 10\msun$ black
hole is met by the expanding red giant or supergiant envelope. The black
hole accretes some of the matter and transfers enough energy to the remaining
matter to expel it.
The convective envelope has constant entropy and must expand
in order to replace the accreted and expelled matter. The black hole drops
in gravitational potential closer to the
He core of the companion in order to furnish the necessary energy to expel
most of the matter. This whole process happens very rapidly.  (In our case of
hypercritical accretion, an accretion disc will be set up
around the black hole, which cannot immediately accept the matter because of
the high angular momentum of the latter. We assume the vicosity to be
high enough so that angular momentum, but little mass, will be advected
outwards. Observed masses $\sim 7\msun$ of high mass black holes in
transient sources as discussed by Brown, Lee \& Bethe (1999)
substantiate this scenario.)

Hypercritical accretion sets in when the envelope density reaches
$\sim 10^{-9}$ cm$^{-3}$. From eqs. (\ref{eq:5.3}) and (\ref{eq:5.7})
in Bethe \& Brown (1998) we can show that the contribution to the
coefficient of dynamical friction $c_d$ from matter inside the accretion
radius
$R_{ac}$, i.e. from matter which can accrete onto the compact object,
is $\delta c_d=2$. Now the total  coefficient of dynamical friction
is $c_d= 6-8$ (Ruffert 1994; Ruffert \& Arnett 1994).
The remainder of $c_d$ comes chiefly from the wake, at greater distances
than $R_{ac}$. Thus for
\be
c_d \gg 2
\label{eq:6.2}
\ee
we encounter highly nonconservative mass transfer and it is plausible
that common envelope evolution ensues for a wider range of $q$ than
usually thought.
In fact, as we show below, out of the $\sim 20\msun$ hydrogen envelope
of a $30\msun$ companion star only $\sim 3 \msun$ is accepted by the
black hole.

>From the above argumentation we believe that $c_d\gg 2$ strongly favors
common envelope evolution.
(For $c_d=2$ the mass transfer is nearly conservative.)
The companion star should have mass less
than $35-40\msun$, because for higher masses it loses mass in an LBL phase.
It should have somewhat higher mass than the black hole, say $15\msun$ in
our case, so that mass transfer is unstable throughout the transfer
of envelope mass but the final transfer is so rapid that it will overshoot
into the range of stable mass transfer.
As noted, Cyg X-1 is probably included in our interval
of masses favorable for common envelope evolution.

Having established the plausibility of common envelope evolution for
the black hole in the expanding H envelope of an
O-star with ZAMS mass of $15-35\msun$, we now carry out this common
envelope evolution for a typical O-star mass of $20\msun$ following
Bethe \& Brown (1998). The He core is $6\msun$.

We choose the coefficient of dynamical friction $c_d$ to be 6.
In the Bethe \& Brown notation $M_A$ is the compact object mass,
$M_B$ the companion mass, and
\be
Y=M_B a^{-1} .
\label{eq:6.3}
\ee
The initial and final $Y$ are related by
\be
\left(\frac{Y_f}{Y_i}\right)^{1+1/(c_d-1)}
=\frac{2.4 M_{B_i}}{M_{A_i}}
\label{eq:6.4}
\ee
depending only on the initial masses. The ratio of final to initial
black-hole masses is given by
\be
\frac{M_{A_f}}{M_{A_i}}
=\left(\frac{Y_f}{Y_i}\right)^{1/(c_d-1)}
=\left(\frac{Y_f}{Y_i}\right)^{1/5}
=1.3
\ee
so the final black-hole mass is $13\msun$ for an initial $10\msun$.
We find
\be
\frac{a_i}{a_f} =\frac{M_{B_i}}{M_{B_f}}\frac{Y_f}{Y_i}
=12
\label{eq:6.6}
\ee
where $M_{B_f}$ is the companion He core mass of $5.3 \msun$.

\setcounter{equation}{0}
\section{MERGERS AFTER COMMON ENVELOPE}
\label{sec:7}

The most important result of the calculation at the end of
Section \ref{sec:6} is that $a_i/a_f$ is large, of the order of 12.
In (\ref{eq:6.6}), there are two factors;
$Y_f/Y_i$ is fairly large, about 4. $M_{B_i}/M_{B_f}$ is also fairly
large, it is the ratio of the entire mass of the star to its He core
taken to be $(0.3)^{-1}$ by Bethe \& Brown (1998).

Thus the radius of the orbit shrinks by a large factor in the common
envelope. We know the range of the final radius $a_f$: its maximum
is given by the condition that the merger should occur within a Hubble
time. According to (\ref{eq:5.7}), it would be about
$7\times 10^6$ km if the orbit were circular. Eccentricity may raise
this about 30\%,
\be
a_f^{max}\approx 9\times 10^{11} {\rm cm} .
\label{eq:7.1}
\ee

The minimum $a_f$ is some multiple of the radius of the He star which
star B becomes after removal of its hydrogen envelope. The He star radius
may be about $5\times 10^{10}$, and we estimate
\be
a_f^{min}=1.5 - 3 \times 10^{11} {\rm cm} .
\label{eq:7.2}
\ee
So $a_f^{max}/a_f^{min}=3\sim 6$. We assume the distribution of $a$ to be
$da/7a$, so the probability of having $a_f$ in the permitted range is
\be
p=\frac{\ln 3}{7} \sim \frac{\ln 6}{7} = 0.15\sim 0.25 .
\label{eq:7.3}
\ee

The center of the useful range is at about  $5\times 10^{11}$ cm.
Multiplying by $a_i/a_f=12$, we get
\be
a_i \lsim 10^{13}{\rm cm}.
\label{eq:7.4}
\ee
This is too large a distance to permit appreciable X-rays to come from the
black hole. If the two compact stars are to merge by gravitational waves
within a Hubble time, their precursors (a black hole and an O- or B-star)
can presumably not be observed as emitting X-rays.

Conversely, objects like Cygnus X-1 probably will not lead to an observable
gravitational-wave merger. When the O-star expands as a giant and goes into
a common envelope with the black hole, the orbit will contract to the
extent that the black hole falls into the He core of the giant; there is
a merger but its gravitational waves will be too low frequency to be observed.
Thus the discussion in our Sections \ref{sec:4} and \ref{sec:5} is
actually irrelevant, given our scenario for common envelope evolution.

Returning to our discussion up to (\ref{eq:7.3}), we have to examine the
values of the initial mass of star B. It should be greater than $15\msun$
to permit common envelope evolution but less than $35\msun$ so as to avoid
undue mass loss by wind. With $M_{A_0}=90\msun$, this means a range of $q$
of 0.2. Multiplying this by $\alpha_1$, in (\ref{eq:3.2}) and by the mean
of $p$ in (\ref{eq:7.3}), the rate of merger of a massive black hole with
another compact star is
\be
\alpha_4 = 4-6\times 10^{-6} {\rm yr}^{-1} {\rm \; per\; galaxy}.
\label{eq:7.5}
\ee

This is in remarkably good agreement with the results for Case H
with hypercritical accretion of Portegies Zwart \& Yungelson, if we
decrease their number by a factor of 5 to take into account our greater
high-mass black-hole mass limit; namely, with this decrease they would
have  $R=7\times 10^{-6}$ yr$^{-1}$ per galaxy.

The great uncertainty is in $\alpha_1$ because we do not know very well
the minimum ZAMS mass which leads to a massive black hole, thus
$\alpha_1$ is likely to be uncertain by a factor 2 either way.
Together with $p$ in (\ref{eq:7.3}), the uncertainty of merger is
about a factor 3, so
\be
\alpha_4= 1 \sim  10\times 10^{-6}\; {\rm yr}^{-1}\;
{\rm per\; galaxy}.
\ee
This rate is much smaller than that of mergers of small black holes with
neutron stars, $\alpha_s\approx 10^{-4}$, as one might expect from the
difficulty of forming massive black holes. But the signal to noise ratio
in the gravitational wave detector depends on the chirp mass as
$(M_{chirp})^{2.5}$. The chirp mass is
\be
M_{chirp} =\mu^{0.6}M^{0.4},
\ee
where $\mu$ is the effective mass and $M$ is the total system mass.
We have
$M_{chirp}\sim 3.3\msun$ for our high-mass black-hole, neutron-star binaries,
as compared with $\sim 1.6 \msun$ for the low-mass black-hole, neutron-star
binaries.
 Thus,
effectively the high-mass black-hole merging rate should be multiplied
by a factor of $\sim 6$.
The two types can be distinguished by the chirp mass.

Star B, after its supernova event, will be either a neutron star or
a small black hole of mass slightly over $1.5\msun$, it has no occasion
to accrete extra mass. Thus, we estimate that signals from massive black
hole mergers with neutron stars should have a frequency about $1/3$ of
those from smaller black holes. Taking the central value of
$5\times 10^{-6}$ from (\ref{eq:7.5}) and multiplying it by 6, we obtain
an effective increase of $\sim 30\%$ over the merging rate
of $10^{-4}$ yr$^{-1}$ for the low-mass black-hole, neutron-star binaries.

\setcounter{equation}{0}
\section{OTHER EFFECTS OF HYDROGEN ``CLOTHING"}
\label{sec:8}

We have seen that the He core of stars evolves quite
differently according as it is ``clothed" with an H envelope
or not. We believe that this difference may explain problems
in some recent investigations.

\subsection{Formation of black holes with light companions}

Portegies Zwart, Verbunt \& Ergma (1997)
have discussed the formation of black holes in low-mass X-ray
binaries. Many of these transient X-ray sources have been discussed
recently with black-hole masses in the probable range $\sim 6-7\msun$.
Assuming a lower limit of $\sim 40\msun$ for black-hole formation,
Portegies Zwart et al. find a much too low rate of formation.

In their evolution, the black hole originates from the more massive
component (A) of the binary. The less massive component (B) will spiral in
to A when A becomes a giant and touches B.
They show that B will survive this spiral-in only if its original
distance from A is at least several hundred solar radii.
Otherwise, B will spiral into
the He core of A, will merge with A and thus be lost. In order for A
to have a radius of several hundred $R_\odot$, it must have completed
He core burning. This means that A has  burned He while
``clothed" with most of its H envelope.
Consequently, the upper curve in Fig.~1 applies; the massive
star A essentially burns as a single star. In this case stars of ZAMS
mass as low as $20-25\msun$ can go into a black hole.
We suggest this as a possible solution to the Portegies Zwart
et al. underproduction of HMBH's.

Our scenario also  suggests an explanation of why the transient
black holes are generally accompanied by low-mass companions.
As the massive progenitor A of the black hole evolves as a giant,
RLOF will generally transfer its
H envelope to a massive companion (if one exists) during hydrogen shell
burning or early in the He core burning phase.
After RLOF the He
core of the primary will burn as a ``naked" He core, with
possible fate as either low-mass black hole or neutron star
(Brown, Weingartner \& Wijers, 1996) but not as a high-mass black hole.
Therefore a high-mass black hole will generally not have a massive companion,
except in the relatively rare cases.

If, on the other hand, star B has small mass, it can accept only very
little mass from A in RLOF. Instead, interaction
of the two stars will wait until A becomes a supergiant and its surface
reaches B. Then, in the common envelope, B spirals in as described
by Portegies Zwart et al. The He core of star A, mass
about 1/3 of the original ZAMS mass, evolves as described by
Portegies Zwart et al., finally collapses as a SN and leaves a remnant
of mass approximately equal to the He core mass, $6-7\msun$.

\subsection{Cygnus X-3}

Cyg X-3 may be a progenitor of a binary of two low-mass black holes.
We believe the envelope of the O-star progenitor of the compact object
(A) in Cyg X-3 was probably lifted off in RLOF. This star A then
had a SN event making it into a neutron star. When star B, the present He
star, become a giant, star A would spiral in which would have converted
the neutron star into a low-mass black hole (if it wasn't one already).
The He star B in Cyg X-3 is certainly burning as ``naked". Our
estimated is that the ZAMS mass of the progenitor of this He star
is about the maximum mass that does not go into an LBL stage (because the
envelope must have been used up in common envelope evolution), say $35\msun$.
>From our fig. 1 the most likely fate of such a star, with H envelope lifted off
in RLOF, is a LMBH, but a neutron star of mass $\sim 1.5\msun$ cannot be
excluded.

\subsection{Nucleosynthesis}

The galactic ratio of oxygen to iron depends on
the ZAMS mass above which single stars evolve into
high mass black holes (cut-off mass)
and therefore do not return matter to the galaxy.
Oxygen is chiefly produced in quiescent burning
before the supernova explosion and the amount is roughly proportional
to ZAMS mass. Fe, on the other hand, is produced explosively
in the SN explosion, and the amount is roughly independent of ZAMS mass,
possibly decreasing slightly as the latter increases. Tsujimoto
et al. (1997) have recently used the [O/Fe] ratio
and observations in metal poor stars to determine a cutoff mass of
\be
M_{{\rm cutoff}} = 50 \pm 10 \msun.
\label{eq:8.1}
\ee

They remark that the influence of the metallicity dependence of stellar
wind losses may be significant, but do not take it into account.
We believe, however, the larger wind losses of stars with solar metallicity
to have important effects. Over a wide range of ZAMS masses, between
$\sim 35-40\msun$ and $\sim 80\msun$ according to Woosley, Langer \& Weaver
(1993), the winds remove the H envelope sufficiently rapidly that the
He cores evolve as ``naked". Thus, the cores evolve into either low-mass
black holes or neutron stars, their (gravitational) core masses following
the lower line heading slightly above $1.5\msun$ in Fig.~1.
Brown, Weingartner \& Wijers found that the primary in 1700-37
probably evolved into a low-mass black hole (I.e., it doesn't pulse.)
whereas 1223-62 is known to contain a neutron star. In both cases the
ZAMS mass of the primary was found to be $\sim 40\msun$, so this
approximately locates the ZAMS mass corresponding to $M_{NP}$ of
Section~\ref{sec:2}, in the case of solar metallicity.
(As noted earlier, we estimate the ZAMS mass corresponding to $M_{PC}$
to be $\sim 80\msun$.) It may well be that for metal poor stars, with much
weaker winds, the hydrogen envelope is not removed rapidly enough
for their He cores to evolve as ``naked". In this case the limit
(\ref{eq:8.1}) for ZAMS masses giving core mass $M_{PC}$ could be more
appropriate.

We have not attempted to carry out a calculation of nucleosynthesis
in our scenario, but wish to point out features of our scenario
which will tend to increase the $M_{{\rm cutoff}}$ above which
element production for the galaxy ceases.

Disregarding mass loss, single stars above a certain mass $M_{min}$
evolve into high-mass black holes. From the upper curve in Fig.~\ref{fig1},
one might take the minimum He mass to be $6-8\; M_\odot$ corresponding to
a ZAMS mass $20-25\; M_\odot$. But as the mass of the single star increases,
there is big mass loss before the star ever reaches the supernova stage.
So with extensive mass loss WLW(1993) find that ZAMS mass $35\; M_\odot$
leads, after SN, to a compact object of about $1.5\; M_\odot$, thus a
low-mass black hole. Thus high-mass black holes are only formed by single
stars in a limited mass range, from $20-25\; M_\odot$ to about
$35 -40 \; M_\odot$.
Stars in this intermediate mass range, of width $10-20\; M_\odot$, do not
return their matter, especially their Fe, to the galaxy.
This group of stars is in addition to the stars above $M_{cutoff}$.
To compensate for this fact, $M_{cutoff}$ will have to be raised
above the Tsujumoto value.
Because the abundance of stars decreases with increasing mass, the raise
has to be more than $10-20\; M_\odot$. Thus we suggest that
$M_{cutoff}$ may be as high as the $80\; M_\odot$ which we used for
other reasons in section~\ref{sec:3}.

We note from the WLW (1993) calculations that
``naked" He cores evolve into less massive carbon/oxygen cores
than ``clothed" ones. Thus we see that inclusion of mass
loss will tend to move
the mass above which all nucleosynthesis ceases even higher.
Therefore, we believe that
inclusion of the mentioned effects may decrease the apparent
discrepancy between $M_{{\rm cutoff}}$ of Eq.~(\ref{eq:8.1}) and
our ZAMS mass of $80\msun$ for making MBH's.

We return to the higher observed incidence of HMBH's (LMC X-1 \&
LMC X-3) in the LMC. Suppose that the limit of Eq.~(\ref{eq:8.1})
applies for the lower metallicity of the LMC; i.e., stars
with ZAMS masses $\sim 40-80\msun$ in the LMC would not experience as
large wind
loss as those in the disc, because of the lower metallicity in the LMC.
This could help to explain why two HMXBs containing high-mass black holes
are observed in the LMC, and only Cyg X-1 is seen in the Galactic disc.

\setcounter{equation}{0}
\section{CONCLUSIONS}
\label{sec:9}

Our chief result is that those massive binary systems in which the primary
is sufficiently massive $\gsim 80\msun$ to go into a high-mass black hole can
contribute importantly to observable gravitational waves upon merger of the
final binary of compact objects. These mergers come chiefly from companion
O-stars in the range of $15-35\msun$ ZAMS mass.

We argue that the situation created by hypercritical accretion is favorable
to common envelope evolution in this range of companion masses, and that
the time scale of this evolution is very fast.

We show that population syntheses do not have a common envelope with
efficiency comparable with ours, possibly because they generally
do not include hypercritical accretion. Certainly they do not have as
effective a tightening of orbits as we have.
The exception is Case H of Portegies Zwart \& Yungelson (1998)
which does include hypercritical accretion. For the same mass limit
for black hole, their results are in good agreement with ours.  At first sight
it may seem surprising that we match so well the Portegies Zwart \& Yungelson
work, provided they include hypercritical accretion as we do.  Their
calculation includes various phases of mass transfer before forming the binary
consisting of a neutron star and black hole.  However, results depend only on
the ratio of the  logarithmic interval favorable for gravitational merger with
avoidance of coalescence to the total logarithmic interval over which binaries
are distributed.  Whereas the favorable logarithmic interval is shifted around
in the various mass transfers, its magnitude is unchanged because the only
scale is the radial separation of the two objects.

Our final merger rate of $\sim 4-6\times 10^{-6}$ yr$^{-1}$ increases
the Bethe \& Brown (1998) rate of gravitational waves by a factor of
$\sim 1.3$, largely because of the higher chirp mass with high-mass
black holes.

This important effect comes in spite of much higher mass limit ZAMS
$80\msun$ for evolution into high-mass black holes in binaries.
We justify this high mass limit from results for the evolution
of ``naked" He stars by Woosley, Langer \& Weaver. Namely, the more
massive primary in binaries has its mass lifted off early, either
RLOF or in LBL stage for the stars with ZAMS masses $> 35 -40\msun$.
The resulting ``naked" Wolf-Rayet or He star is
deprived of the H-envelope which normally ``insulates" He cores during
their burning so the convective carbon burning stage is not skipped
as it is in single stars of ZAMS mass $\gsim 20\msun$. The great
entropy loss during the long duration of this stage results in a low-mass
compact object.

We show that the high mass limit for evolution into high-mass black holes
in binaries is consistent with nucleosynthesis, because single stars
with ZAMS masses in the range of 20 to $35-40\msun$ evolve into high-mass
black holes without return of matter to the galaxy.

We also indicate that we could evolve enough transient sources, binaries
of a high-mass black hole and low mass main sequence star.

There is interest in the merging of binaries composed of two
high-mass black holes (Brady, Creighton \& Thorne 1998).
With our mass limit of $80\msun$ for high-mass black holes the
initial separation of the two massive O-star progenitors must be
$>40R_\odot$ for both of them to lie inside their Roche Lobes.
Case A mass transfer between two nearly equal mass stars can only
widen the binary, since $a_f/a_i=(\mu_i/\mu_f)^2$ and $\mu$ will
decrease in the transfer. Stars of ZAMS mass $\gsim 40\msun$ lose
their H envelopes rapidly in the LBL phase, so a common envelope
phase in which the orbit is tightened is excluded. The Wolf-Rayets
which result after loss of the H will lose most of their mass by wind,
down to $\sim 10-15\msun$ before going into black holes. With
spherically symmetric mass loss, $a_f/a_i=M_i/M_f\sim 4$.
Thus, the separation of the two final black holes will be several times
the initial $>40~{\rm R}_\odot$ which precludes merging in a Hubble time.
We are in agreement with Portegies Zwart \& Yungelson (1998) on this point.

We believe that in Brown \& Bethe (1998) together with this paper we
have given a consistent description  of the evolution
of most binaries of compact stars containing a black hole.

\vskip 2cm


We would like to thank the referees for extremely helpful criticism.
We are grateful to Simon Portegies Zwart and Ralph Wijers for advice.
We would like to thank Chang-Hwan Lee for useful advice and help, and
Fred Walter for information on observability.
We are grateful to Kip Thorne for discussion of the signal to noise
ratio in LIGO.
G.E.B. was partially supported by the U.S. Department of Energy under Grant No.
DE--FG02--88ER40388.

\centerline {REFERENCES}
\begin{tabbing}
xxxxx\=xxxxxxxxxxxxxxxxxxxxxxxxxxxxxxxxxxxxxxxxxxxxxxxxxxxxxxxxxxxxxxxxxxxx\kill

\ni Bethe, H. A. \& Brown, G. E. 1998, ApJ, 506, 780. \\
\ni Brady, P.R., Creighton, J.D.E., \& Thorne, K.S. 1998,
Phys. Rev. D 58, 061501.\\
\ni Brown, G. E. 1997, Phys. Bl. 53, 671.\\
\ni Brown, G. E. \& Bethe, H. A. 1994, ApJ, 423, 659. \\
\ni Brown, G. E., Lee, C.-H., \& Bethe, H. A. 1999, submitted to
 Phys. Repts.\\
\ni Brown, G. E., Weingartner, J. C., \& Wijers, R. A. M. J. 1996,
  ApJ, 463, 297. \\
\ni Cappelaro, E., Turatto, M., Tsuetkov, D.Yu.,
Bartumov, O.S., Polls, C., Evans, R., \\
 \> \& Hasnuy, M. 1997, A\&A, 322, 431. \\
\ni Cordes, J.M., \& Chernoff, D.F. 1997, ApJ, 482, 971.\\
\ni Garmany, C.D., Conti, P.S., \& Massey, P. 1980, ApJ, 242, 1063.\\
\ni Herrero, A., Kudritzki, R.P., Gabler, R., Vilchez, J.M.,
and Gabler, A. 1995, A\&A 297, 556.\\
\ni Iben, I., Tutukov, A. V., \& Yungelson, L. R. 1995, Ap \& SS, 100, 217. \\
\ni Kalogera, V. 1996, ApJ 471, 352. \\
\ni Kippenhahn, R., \& Meyer-Hofmeister, E. 1977, A\&A 54, 539.\\
\ni Kuiper, L., van Paradijs, J., \& van den Klis 1997, A\&A, 79, 203.\\
\ni Li, G.Q., Lee, C.-H., \& Brown, G.E. 1997, Nucl. Phys. A625, 372;\\
  \> Phys. Rev. Lett. 79, 5214.\\
\ni Lipunov, V.M., Postnov, K.A., \& Prokhorov, M.E. 1997, MNRAS,
288, 245.\\
\ni Massevich, A. G., Popova, E. I., Tutukov, A. V., \& Yungelson, L. R. 1979,
   \\ \> Ap \& SS, 62, 451. \\
\ni Moffat, A.F.J. \& Carmelle, R. 1994, ApJ, 421, 310. \\
\ni Phinney, E.S. 1991, ApJ, 380, L17.\\
\ni Poisson, E., \& Will, C. M. 1995, Phys. Rev. D, 52, 848. \\
\ni Portegies Zwart, S. F. \& Yungelson, L. R. 1998, A\&A, 332, 173. \\
\ni Portegies Zwart, S. F., Verbunt, F., \& Ergma, E.
  1997, A\&A, 321, 207.\\
\ni Prakash, M., Bombacci, I., Prakash, M., Ellis, P.J., Lattimer, J.M.,
    \& Knorren, R. 1997,
   \\ \> Phys. Rep., 280, 1. \\
\ni Rasio, F.A., \& Livio, M. 1996, ApJ, 471, 366.\\
\ni Ruffert, M. 1994, ApJ, 427, 351.\\
\ni Ruffert, M. \& Arnett, D. 1994, A\&AS, 106, 505.\\
\ni St-Louis, N. et al. 1993, apJ 410, 342. \\
\ni Terman, J.L., Taam, R.E., \& Hernquist, L. 1995, ApJ, 445, 367;
 see especially fig. 3.\\
\ni Thorne, K. 1998, private communication.\\
\ni Thorsett, S.E., \& Chakrabarty, D. 1998, astro-ph/9803260, ApJ submitted.\\
\ni Thorsson, V., Prakash, M., \& Lattimer, J.M. 1994, Nucl. Phys. A572, 693.\\
\ni Tsujimoto, T., Yoshii, Y., Nomoto, K., Matteuci, F.,
Thielemann, F.-K. \\ \> \& Hashimoto, M. 1997, ApJ, 483, 228.\\
\ni Van den Heuvel, E. P. J., \& Lorimer, D. R. 1996, MNRAS, 283, L37. \\
\ni Van den Heuvel, E. P. J 1994 (Saas Fe Advances Course 22,
 Springer Verlag, Berlin) \\ \> eds. H. Nussbaumer \& A. Orr, p. 263.\\
\ni Woosley, S. E., Langer, N., \& Weaver, T. A. 1993, ApJ, 411, 823. \\
\ni -----, 1995, ApJ, 448, 315. \\
\ni Woosley, S. E., \& Weaver, T. A. 1995, ApJS, 101, 181.
\end{tabbing}

\newpage
\centerline{FIGURE CAPTION}
\noindent FIG. 1-- Comparison of the compact core masses resulting from the
evolution of single stars (filled symbols), Case B of Woosley \& Weaver (1995),
and naked helium stars (Woosley, Langer \& Weaver, 1995) with masses equal to
the corresponding He core mass of single stars.  The horizontal dashed lines
indicate the mass of the heaviest known well--measured pulsar and the maximum
mass of a neutron star.
\newpage

\end{document}